\documentclass[twocolumn,showpacs,preprintnumbers,amsmath,amssymb]{revtex4}
\usepackage{epsf}

\begin{document}

\preprint{Theoretical Framework for 
 Microscopic Osmotic Phenomena}

\title{Theoretical Framework for Microscopic Osmotic Phenomena}

\author{Paul J. Atzberger}
\thanks{\textit{address:} University of California; 
Department of Mathematics; Santa Barbara, CA 93106;
\textit{e-mail:} atzberg@math.ucsb.edu; \textit{phone:} 805 - 679 - 1330.}
\affiliation{Department of Mathematics \\ 
University of California at Santa Barbara.}

\author{Peter R. Kramer}
\affiliation{Department of Mathematical Sciences \\ 
Rensselaer Polytechnic Institute.}

\date{\today}

\begin{abstract}
The basic ingredients of osmotic pressure are a solvent fluid with a 
soluble molecular species which 
is restricted to a chamber by a boundary which is 
permeable to the solvent fluid but impermeable to the solute
molecules.  For macroscopic systems at equilibrium,
the osmotic pressure is given by the classical 
van't Hoff Law, which states that the pressure 
is proportional to the product of the temperature 
and the difference of the solute concentrations 
inside and outside the chamber.  For microscopic systems
the diameter of the chamber may be comparable to the 
length-scale associated with the solute-wall interactions 
or solute molecular interactions.  In each of
these cases, the assumptions underlying the 
classical van't Hoff Law may no longer hold.
In this paper we develop a general theoretical framework 
 which captures 
corrections to the classical theory for 
 the osmotic 
pressure under more general relationships between the size 
of the chamber and the interaction length scales.  We also 
show that notions of osmotic pressure based on the hydrostatic 
pressure of the fluid and the mechanical pressure on the 
bounding walls of the chamber must be 
distinguished for microscopic systems.  To demonstrate 
how the theoretical framework 
 can be applied, numerical results are 
presented for the osmotic pressure associated with a 
polymer of $N$ monomers confined in a spherical chamber 
as the bond strength is varied.
\end{abstract}

\maketitle

\section{\label{sec:intro}Introduction}
Osmotic effects are thought to play an important 
role in many physical systems associated with 
biology and technological applications. 
Examples in technological applications
include microscopic devices 
designed to pump fluids~\citep{atzbergerPump2006,bazant2004,nardi1999},
actuate forces through swelling~\citep{su2002,brenner1996},
or deliver drug doses~\citep{theeuwes1976,verma2002,wolgemutha_hydra_2004}.  
Some macroscopic osmotic mechanisms in biology 
include the exchange of blood constituents in 
capillaries with surrounding tissues~\citep{hammel1999}  
and the processing of fluids in tissues of layered
epithelial cells in the intestines and 
kidneys~\citep{duquette2001,spring1999}.  

At a more microscopic level, individual cells
contain a high concentration of charged 
molecules, in which osmotic effects must 
be controlled actively to avoid excessive 
swelling and bursting of cellular structures
\citep{hoppensteadt2001,go:doff}.
In fact some of the mechanisms by which neural cells 
transmit electrical signals through the production 
and propagation of action potentials may have as 
their evolutionary origins the pumping mechanisms 
developed by cells to use counter-ion fluxes to 
compensate for the harmful effects of osmotic pressure 
\citep{douzou1994,hoppensteadt2001}.
Other examples include the study of the pressures involved in DNA 
confinement in virus capsids~\citep{evilevitch2003}, 
packaging of proteins in small cellular vesicles in cell organelles~\citep{thecell}, 
and even mechanisms of gel swelling in propulsion in micro-organisms such 
as myxobacteria~\citep{wolgemutha_myx_2004,wolgemutha_hydra_2004,nardi1999}.

With new experimental techniques, such as optical trapping 
and molecular tagging, it is now feasible to observe and 
measure forces and displacements in systems on the 
length-scale of hundreds to tens of nanometers~\citep{block2003,lippincott2003,danuser2003}, 
and the underlying physical processes can begin to be explored quantitatively at 
very small length-scales in biological and 
synthetic systems~\citep{block2003,danuser2003,lipowsky1999,evilevitch2003,levich1962,einstein1926}.
The modeling and analysis of these types of systems motivates using 
a theoretical framework from statistical mechanics which is general 
enough to apply to these systems without the usual type of 
hard-wall and other scale separation assumptions appropriate 
in classical thermodynamic systems.

In this paper we shall discuss a general microscopic theory for 
osmotic pressure at equilibrium and draw some contrasts 
with the classical theory of van't Hoff~\citep{vanthoff1887}.  
First, in Section \ref{sec:microscopic_model}, we introduce  two 
statistical quantities to characterize pressure associated with 
osmotic phenomena: one 
related to the pressure built up in the fluid and another to 
the pressure felt by the confining wall of the chamber.
We find in Section \ref{sec:stat_mech_formulas}  
that while both lead macroscopically to equivalent notions
of osmotic pressure, they differ in general for microscopic 
systems with interaction length scales comparable to the 
chamber size.  Each form
of osmotic pressure has theoretical connections to macroscopic thermodynamics:
the osmotic wall pressure maintains its statistical mechanical relationship to changes in free energy with respect to volume changes, while the osmotic fluid pressure is directly related to the notion of osmotic pressure in macroscale nonequilibrium thermodynamics. 
A specific example 
illustrating the distinctions between the two notions of osmotic 
pressure and their deviations from the classical van't Hoff law is 
presented in Section \ref{sec:steric_interaction} and 
Section \ref{sec:softwall_interaction}, 
in which noninteracting solutes have a solute-wall
interaction potential given by a power-law of the 
distance of the solute to the wall.

We then proceed in Section 
\ref{sec:interact_string} 
with an example 
 with solute particles bound by 
``string" 
forces in a hard-walled potential 
to illustrate how 
the interaction length scale between solute particles affects 
the osmotic pressure.     After preparing in Section~\ref{sec:interact_n_gen} with an alternative expression for the
osmotic pressure in terms of corrections from the van't Hoff law, 
we illustrate in Section~\ref{sec:confined_polymers} the theoretical formalism through numerical Monte Carlo calculations for the osmotic pressure for a polymer confined within a spherical chamber with various binding strengths (and therefore bond lengths).  The theoretical framework 
 presented in this paper is expected to 
be applicable to model osmotic effects in many microscopic systems at 
equilibrium, and 
provides a step toward developing a 
non-equilibrium (or near-equilibrium) theory for microscopic osmotic 
phenomena.

\section{\label{sec:microscopic_model}
Microscopic Osmotic Pressure} 
We begin by posing two possible means for describing the pressures 
associated with osmotic phenomena at a microscopic level in a 
chamber with a confining boundary (wall) 
permeable to the solvent but not the solute.
We are concerned here with developing a microscopic analysis,
rather than a macroscale thermodynamic description.  In particular,
we are interested in a theory based on quantities that could  be 
measured in a microscale experiment or computed from a microscale 
simulation, using a numerical method for example such as the stochastic 
immersed boundary method \citep{atzberger2005ib} or Stokesian dynamics 
\citep{jfb:sd}.  

The
first quantity we shall consider is the mechanical pressure exerted 
by the solute on the wall. The second quantity is the hydrostatic 
pressure built up in the solvent in the interior of the chamber.  
While these two notions lead to the same pressure quantities for 
macroscopic systems, referred to as the osmotic pressure,
we shall show that different results are obtained for the 
pressures when considering microscopic systems.  

The osmotic pressure definition based on the wall pressure 
maintains some classical thermodynamic relationships involving pressure 
and other statistical mechanical quantities in nonideal systems for which 
the length scale of the wall and solute interactions are nonzero and finite.  
On the other hand, the osmotic pressure definition based on the manifested 
fluid pressure can be shown to be equivalent, in a certain local 
sense, to the osmotic pressure definition used in the theory of macroscopic 
nonequilibrium thermodynamics 
\citep{katchalsky1965,IBler:mcsp}
to describe 
the driving force behind solute fluxes across a semipermeable wall or membrane.  

Both notions of osmotic pressure therefore have some underlying theoretical 
justification in both microscopic terms and in relation to macroscopic 
thermodynamics, though we will show through theory and example that they 
are not equivalent.  Both types of pressure moreover would appear to have 
practical relevance for modeling and analysis depending on the application.

When investigating swelling phenomena and the elastic forces of 
a confining membrane, for example, a notion of pressure involving 
the average force exerted per unit area on the confining wall may 
be of particular interest.  In contrast, when investigating 
the role of fluid dynamics in transport and dilution of the 
concentration of solute, for example,  the hydrodynamic pressures 
which drive the flows may be of more relevance.

In what follows,
we generally restrict attention to the case of $ N$ identical 
and possibly interacting solute particles 
confined in the chamber.   
Furthermore, we shall assume 
throughout that the potential energy of the particles is of the form: 
\begin{eqnarray}
\Psi(\mathbf{x}_1,\cdots,\mathbf{x}_N)
& = & 
V(\mathbf{x}_1,\cdots,\mathbf{x}_N)
+
\sum_{k = 1}^N 
\Phi(\mathbf{x}_k).
\end{eqnarray}
In the notation,
$V(\mathbf{x}_1,\cdots,\mathbf{x}_N)$ models the energy of the solute-solute interactions
and $\Phi(\mathbf{x})$ models the interaction of a solute particle with the chamber wall.
The $ \mathbf{x}_k $ denote the spatial coordinates for the $k$th solute particle. 
Much of the theory can be readily extended to more general potentials.

\subsection{\label{sec:wall_pressure}
Osmotic Wall Pressure}
The first notion of ``osmotic pressure"
we consider will be defined in terms of the 
average forces that solute particles exert on 
the walls of a confining chamber.  We shall assume that 
the solute-wall interaction potential $\Phi$ arises from a 
uniform areal distribution of particles on the wall which 
have an isotropic interaction with the solute, given by a potential
$\phi$. More specifically,
we shall assume that for a confined solute particle at interior 
location $ \mathbf{x} $ the interaction force exerted on the 
differential area $ d \mathbf{y} $ of  
the bounding wall can be expressed as:
\begin{eqnarray}
\mathbf{G}(\mathbf{x} - \mathbf{y}) d \mathbf{y} 
\end{eqnarray}
where $\mathbf{G} (\mathbf{r}) = -\nabla_{\mathbf{r}} \phi(|\mathbf{r}|)$ 
and $\phi$ is 
the isotropic interaction potential of 
the wall particles with the solute particles.

The force acting on a solute particle in the chamber interior which 
arises from the wall interactions is then given by:
\begin{eqnarray}
\mathbf{F}_{\text{wl}} 
(\mathbf{x}) 
& = & \int_{\partial \Omega} -\mathbf{G}(\mathbf{x} - \mathbf{y}) d\mathbf{y} 
\label{eq:FG} 
\end{eqnarray}
where $ \Omega $ defines the container volume and $ \partial \Omega $ the 
container boundary, which we often simply call ``a wall."  We discuss in Appendix~\ref{sec:inversion} how the form of the solute-wall interaction ($ \mathbf{G} $ or $ \phi $) can be inferred from an observed or known structure for the wall force $ \mathbf{F}_{\text{wl}} $ or potential $ \Phi $.  

To keep focus on the main issues of interest, we will treat the 
container shape as prescribed and fixed.  The results would apply 
to more general systems with a flexible boundary (such as a 
membrane-bound vesicle), provided the thermal fluctuations of 
the bounding surface in equilibrium could be neglected due to 
their amplitude or time scale.  In this case, we would not be 
predicting self-consistently the shape of the container, but 
simply taking its (time-averaged) shape in thermal equilibrium 
as an input to define $ \Omega $.

Forces of the form Eq. (\ref{eq:FG})
can arise 
in a system in a variety of ways.
For example, a solute particle (possibly a macro-ion) 
in the interior of a chamber could interact with 
like-charged particles composing the wall surface~\citep{grosberg2002}.
In this case, $\mathbf{G}$ would represent the 
screened Coloumbic force per unit area 
exerted on the surface as a consequence 
of the charge density of the wall and solute
while $\mathbf{F}_{\text{wl}}$ would be the screened Coloumbic 
force acting on the solute particle when it occupies 
location $\mathbf{x}$.  In principle, both of these forces 
could be derived from a common electrostatic potential for 
the system~\citep{prabhu2005}.  Another example would
be for $\mathbf{G}$ to represent an effective force
of interaction between a particle and the wall
arising from time-averaged steric interaction forces 
such as those between a particle and a polymer brush
coating the wall~\citep{auroy1992}.  One could even 
consider within this framework  the degrees of freedom of a single 
monomer of a large, compactly folded polymer as constituting an
effective particle confined by the time-averaged effects of the rest
of the monomers in the polymer. 

To define a pressure for these types of 
systems we can consider the average forces 
which act in the normal direction across 
the wall surface:
\begin{eqnarray}
P_{\mathrm{wl}} & = & 
\left\langle
\frac{1}{|\partial \Omega|}\int_{\partial \Omega} 
\sum_{j=1}^n \mathbf{G}(\mathbf{X}^{[j]}(t) - \mathbf{y})\cdot 
\hat{\mathbf{n}}_\mathbf{y} \, d\mathbf{y}
\right\rangle 
\label{P_wl_definition}
\end{eqnarray}
where $|\partial \Omega|$ denotes the surface area of bounding 
wall $\partial \Omega$, $ \hat{\mathbf{n}}_{\mathbf{y}} $ 
denotes the unit outward normal to the surface at 
$ \mathbf{y} $ and $\langle \cdot \rangle$ denotes a 
time average over the position of the $N$ solute particles 
$\{\mathbf{X}^{[j]}(t)\}_{j=1}^N$, which we shall
assume is equal to the statistical ensemble average.
In the notation,
$ \mathbf{X}^{[j]} (t) $ denotes the position 
of solute particle $j$ at time $t$.
We shall refer to this definition of pressure as the 
``osmotic wall pressure."

We can alternatively express this osmotic wall pressure 
in terms of the density of the wall-solute interaction potential as:
\begin{eqnarray}
\nonumber
P_{\mathrm{wl}} & = & 
\left\langle
\frac{1}{|\partial \Omega|}\int_{\partial \Omega} 
\sum_{j=1}^n - \hat{\mathbf{n}}_\mathbf{y} 
\cdot \nabla_{\mathbf{y}} \phi(|\mathbf{X}^{[j]}(t) - \mathbf{y}|)
\, d\mathbf{y}
\right\rangle \\
\nonumber
& = &
\left\langle
\frac{1}{|\partial \Omega|}\int_{\partial \Omega} \int_{\Omega} 
\sum_{j=1}^n - \hat{\mathbf{n}}_\mathbf{y} \cdot \nabla_{\mathbf{y}} \phi(|\mathbf{x} - \mathbf{y}|)
\cdot
\right.
\\
&& 
\nonumber
\left.
\hspace{3cm}
\cdot
\delta (\mathbf{x} - \mathbf{X}^{[j]} (t))\, d \mathbf{x} \, d\mathbf{y}
\right\rangle  \\
& = & \frac{1}{|\partial \Omega|}\int_{\partial \Omega} \int_{\Omega} 
 - \hat{\mathbf{n}}_\mathbf{y} \cdot \nabla_{\mathbf{y}} \phi(|\mathbf{x} - \mathbf{y}|)
 c (\mathbf{x}) \, d \mathbf{x} \, d\mathbf{y} 
 \label{P_wl_conc}
\end{eqnarray}
where we have introduced the Dirac delta function $ \delta (\mathbf{x}) $ in order to express the statistical average in terms of the solute concentration 
~\citep{IBrk:sp2}:
\begin{eqnarray}
\label{def_conc1}
c(\mathbf{x}) := \left\langle
\sum_{j=1}^n \delta \left(\mathbf{x}  - \mathbf{X}^{[j]}(t)\right)
\right\rangle.
\end{eqnarray}
This concentration is proportional to the probability density
for any of the particle positions $\mathbf{X}^{[j]}(t)$ to be at
location $\mathbf{x}$ when the system is in
statistical equilibrium.

\subsection{\label{sec:fluid_pressure}
Osmotic Fluid Pressure}
Another approach to studying osmotic pressure is to 
consider the solvent fluid and, in particular, how the
forces of interaction between the solute and wall 
are manifested in the hydrostatic pressure. 
For a Stokesian fluid we have that the local fluid 
velocity $ \mathbf{u} $ satisfies: 
\begin{eqnarray}
\rho \frac{\partial \mathbf{u}}{\partial t} 
&=& -\mu \Delta{\mathbf{u}} -\nabla{p} 
+ \mathbf{f} + \mathbf{f}_{\text{th}} \\
\nabla \cdot \mathbf{u} & = & 0 
\end{eqnarray}
where $ \rho $ is the fluid density, $ \mu $ is 
the dynamic viscosity, $p $ is the fluid pressure, 
$\mathbf{f}  $ is the force density 
exerted by the solute particles, $ \mathbf{f}_{\text{th}} $ 
is the force density associated to thermal fluctuations 
\citep{atzberger2005ib,IBldl:ctp6,IBldl:ctp9}.

Working with the immersed boundary method
approximation~\citep{peskin2002} 
for the interaction of a 
fluid with $ N $ identical solute particles, we 
express: 
\begin{eqnarray}
\mathbf{f} (\mathbf{x},t) & = & \sum_{j = 1}^N
-\nabla_{j}\Psi (\mathbf{\Lambda}^j) \delta \left(\mathbf{x} 
- \mathbf{X}^{[j]} (t)\right).
\end{eqnarray}
In the notation,
$ \mathbf{X}^{[j]} (t) $ denotes the position 
of solute particle $j$ at time $t$ and 
$ \nabla_j $ denotes 
a gradient with respect to the spatial coordinate of solute 
particle $ j $.  For compactness in the notation we 
also define convenient  
expressions involving composite
vectors of the particle positions as follows:
$\mathbf{{\Lambda}}^{j} = \mathbf{{\Lambda}}^{j}(\mathbf{x},\{\mathbf{X}^{[k]} (t)\}_{k = 1, k \neq j}^N) = \left[\mathbf{X}^{[1]},\ldots ,\mathbf{X}^{[j-1]},\mathbf{x},\mathbf{X}^{[j+1]},\ldots ,\mathbf{X}^{[N]}\right]^T$
and
$\mathbf{\Lambda}_{j} = \mathbf{\Lambda}_{j}(\mathbf{x},\{\mathbf{x}_k\}_{k = 1, k \neq j}^N) = \left[\mathbf{x}_{1},
\ldots ,\mathbf{x}_{j-1},\mathbf{x},\mathbf{x}_{j+1},\ldots ,\mathbf{x}_{N}\right]^T$.
The $ \delta (\mathbf{x}) $ 
describes the manner in which the force on the solute particles 
is distributed to the fluid; for small idealized point particles 
this would just be the usual Dirac $\delta$-function.

More 
general particle-fluid interaction functions, however, could be 
used both for modeling and numerical reasons, as in the 
immersed boundary method~\citep{peskin2002}.  
Our end results will generally be posed in a manner which can be appropriate for $\delta $ functions with zero or finite width, but for conciseness we will treat the $ \delta $ functions in derivations as classical Dirac delta functions.  
Moreover, one could more rigorously describe the fluid-particle 
interactions in terms of rigid or flexible moving boundaries, 
but we choose to use the immersed boundary approximation 
with Dirac $\delta $ functions in this initial work since 
it will convey our central ideas and distinctions with a 
minimum of technical distraction.  

Since no external driving force is applied to the chamber 
we have at statistical equilibrium that the average 
fluid velocity must be 
$\left\langle\mathbf{u}(\mathbf{x})\right\rangle = 0$,~\citep{groot1951}, 
which, along with the fact that the thermal force density $ \mathbf{f}_{\text{th}} $ has zero mean, implies 
\begin{eqnarray}
0 & = &
-\left\langle\nabla{p}
\right\rangle + \left\langle\mathbf{f}\right\rangle.
\label{equ_avg_laplacian_u}
\end{eqnarray}
Expressing the force density $ \mathbf{f} $ explicitly in terms of the potentials of the forces acting on the solute particles, we can define an average pressure gradient:
\begin{eqnarray}
\nabla{\bar{p}}(\mathbf{x}) 
& := & \left\langle\nabla{p}(\mathbf{x}) \right\rangle = \left\langle\mathbf{f}(\mathbf{x}) \right\rangle
\nonumber \\
&=& 
 \left\langle \sum_{j=1}^N
-\nabla_{\mathbf{x}}{\Phi(\mathbf{x})} \delta \left(\mathbf{x} 
- \mathbf{X}^{[j]} (t)\right) \right\rangle \nonumber \\
& + & 
\left\langle \sum_{j=1}^N
-\nabla_{j}V(\mathbf{{\Lambda}}^{j})
 \delta \left(\mathbf{x} 
- \mathbf{X}^{[j]} (t)\right) \right\rangle \nonumber \\
&=& 
-\nabla_{\mathbf{x}}{\Phi(\mathbf{x})} c(\mathbf{x}) \nonumber \\
& + & 
\left\langle \sum_{j=1}^N 
-\nabla_{j}V(\mathbf{{\Lambda}}^{j})
 \delta \left(\mathbf{x} 
- \mathbf{X}^{[j]} (t)\right) \cdot \right. \nonumber \\
&&\left. \cdot
\int_{{\Omega^{N - 1}}}
\prod_{k = 1, k\neq j}^{N} \left(\delta\left(\mathbf{x}_k - \mathbf{X}^{[k]}(t)\right) 
d\mathbf{x}_k\right)
\right\rangle \nonumber \\
&=& 
-\nabla_{\mathbf{x}}{\Phi(\mathbf{x})} c(\mathbf{x}) \nonumber \\
& + & 
\sum_{j=1}^N 
\int_{{\Omega^{N - 1}}}
-\nabla_{\mathbf{x}_j}V(\mathbf{{\Lambda}}_{j})\rho_N(\mathbf{{\Lambda}}_{j})
\prod_{k = 1, k\neq j}^{N} d\mathbf{x}_k \nonumber
.\\
\label{def_avg_pressure_grad}
\end{eqnarray}
In the notation, $c(\mathbf{x})$ is the local concentration density 
of the solute as defined in 
Eq. (\ref{def_conc1}) and 
\begin{eqnarray}
\rho_N (\mathbf{x}_{1},\mathbf{x}_{2},\ldots,\mathbf{x}_{N})
= \left\langle 
\prod_{j=1}^{N}
\delta(\mathbf{x}_{j} - \mathbf{X}^{[j]} (t))
\right\rangle
\label{equ_rho_n}
\end{eqnarray}
is the $N$ particle correlation function in thermal equilibrium.  
All terms in the last sum are equal if the particles are truly 
identical and exchangeable, but we leave the expression as an 
explicit sum to incorporate systems with bonded interactions.
For example, in Section \ref{sec:confined_polymers} we consider
a confined polymer in which the monomers on the ends 
have a different bonding structure than monomers in the interior of the 
polymer chain (ends have only one bond, interior monomers have two).

By integrating the pressure gradient along a path 
 starting from a reference point
$ \mathbf{x}_{A} $ outside the domain $ \Omega $ (where we set the pressure to the reference value of zero), we can thereby define what we will call the ``osmotic fluid pressure'' $ P_{\text{fl}} (\mathbf{x}) $ inside the domain:
\begin{eqnarray}
P_{\mathrm{fl}}(\mathbf{x}) & := & \bar{p}(\mathbf{x})
= \int_{\mathbf{x}_A}^{\mathbf{x}} 
\nabla \bar{p} (\mathbf{x}) \cdot d\mathbf{x}.
\label{P_fl_definition}
\end{eqnarray}
To compare this notion of pressure with the osmotic wall
pressure we shall mostly present the osmotic fluid pressure 
at the center of the domain (which we shall take to be 
$ \mathbf{x} = \mathbf{0} $), but 
the definition of the osmotic fluid pressure field extends
throughout space.

\section{\label{sec:stat_mech_formulas}
Statistical Mechanical Formulas for Osmotically Related Pressures}
  
As a starting point for our ensuing analysis, we show how both osmotically related pressures~Eq.~(\ref{P_wl_definition})~and~Eq.~(\ref{P_fl_definition}) defined 
in Section~\ref{sec:microscopic_model} can be represented in terms of the partition function for the solute
particles in a soft-walled potential.
   In Subsection~\ref{sec:hardwall}, we will verify that for noninteracting particles in 
    the hard-wall limit, we recover the standard van't Hoff's law for osmotic pressure both on the wall and in the fluid. 

We consider first the pressure~Eq.~(\ref{P_wl_definition}) induced by the solute 
against the chamber wall.  
In what follows, we will relate the osmotic pressure to changes in the Helmholtz free energy (or equivalently the partition function in the canonical ensemble) under the deformation of the container volume, and the formulas will only take familiar statistical mechanical form if we scale the wall-particle interaction force (and energy) density inversely with the local
 area of the surface.  This corresponds to the physical situation in which the number of particles making up the wall, rather than their areal density, remains fixed under deformation.  Therefore, we define a one-parameter family of 
deformations (diffeomorphisms) $ \boldsymbol{\eta}_s (\cdot)$ 
which deform the chamber boundary $ \partial \Omega \equiv \partial \Omega (s_0) $ into a new boundary $ \partial \Omega (s) $ by extending or contracting the surface at a constant rate along the local outward normal vector $ \hat{\mathbf{n}}_{\mathbf{y}} $ at each $ \mathbf{y} \in \partial \Omega (s)$.  The differential equations describing this family of 
deformations are:
\begin{eqnarray}
\boldsymbol{\eta}_{s_0} (\mathbf{y}) &=& \mathbf{y}, \\
\frac{\partial \boldsymbol{\eta}_s (\mathbf{y})}{\partial s} &=& \hat{\mathbf{n}}_{\boldsymbol{\eta}_s(\mathbf{y})}.
\end{eqnarray} 
These mappings are generally smooth only over a finite interval of parameters $ s $ containing $ s_0 $.  Then, 
taking our $ \phi $ and $ \mathbf{G} $ to refer to the potentials and forces associated with the reference chamber surface $ \partial \Omega (s_0) $, we define the potential for solute-wall interaction forces in 
the deformed chamber boundaries through:
\begin{eqnarray}
\Phi(\mathbf{x},s) & = & \int_{\partial \Omega(s)}  \phi(|\mathbf{x} - \mathbf{y}|) \det (\nabla \boldsymbol{\eta}_s^{-1} (\mathbf{y})) \, d\mathbf{y}, 
\label{def_Phi_int}
\end{eqnarray}
where the Jacobian factor $ \det (\nabla \boldsymbol{\eta}_s^{-1} (\mathbf{y})) $ is inversely proportional to the local expansion of area under the deformation.
  The potential so defined is naturally related to the force defined above in~Eq.~(\ref{eq:FG}) through 
$ \mathbf{F} (\mathbf{x}) = - \nabla_{\mathbf{x}} \Phi(\mathbf{x},s_0)$.  
The total 
normal force acting outward on the surface of the wall exerted by a solute 
molecule at location $\mathbf{x}$ is then given by:
\begin{eqnarray}
h(\mathbf{x}) 
& \equiv &
 \int_{\partial \Omega(s)} -\left[\hat{\mathbf{n}}_{\mathbf{y}} \cdot \nabla_{\mathbf{y}} \phi(|\mathbf{x} - \mathbf{y}|)\right]
\det (\nabla \boldsymbol{\eta}_s^{-1} (\mathbf{y}))\, d \mathbf{y} \nonumber \\
& = & \int_{\partial \Omega(s)} \hat{\mathbf{n}}_{\mathbf{y}} \cdot \nabla_{\mathbf{x}} \phi(|\mathbf{x} - \mathbf{y}|) 
\det (\nabla \boldsymbol{\eta}_s^{-1} (\mathbf{y})) \, d \mathbf{y} \nonumber \\
&=& \int_{\partial \Omega(s_0)} \hat{\mathbf{n}}_{\boldsymbol{\eta}_s (\mathbf{y}^{\prime})} \cdot \nabla_{\mathbf{x}} \phi(|\mathbf{x} - \boldsymbol{\eta}_s(\mathbf{y}^{\prime})|) \,
\, d \mathbf{y}^{\prime} \nonumber \\
&=& \int_{\partial \Omega(s_0)} \frac{\partial \boldsymbol{\eta}_s (\mathbf{y}^\prime)}{\partial s}
\cdot \nabla_{\mathbf{x}} \phi(|\mathbf{x} - \boldsymbol{\eta}_s(\mathbf{y}^{\prime})|) \,
\, d \mathbf{y}^{\prime} \nonumber \\
&=& -\frac{\partial}{\partial s} \int_{\partial \Omega(s_0)} \phi(|\mathbf{x} - \boldsymbol{\eta}_s(\mathbf{y}^{\prime})|) \,
\, d \mathbf{y}^{\prime} \nonumber \\
&=& 
-\frac{\partial}{\partial s}  \int_{\partial \Omega(s)}  \phi(|\mathbf{x} - \mathbf{y}|) 
\det (\nabla \boldsymbol{\eta}_s^{-1} (\mathbf{y})) \, d \mathbf{y}  \nonumber \\
&=& - \frac{\partial \Phi (\mathbf{x},s)}{\partial s}. \label{wall_force}
\end{eqnarray}

The solute concentration $ c( \mathbf{x}) $ is obtained as a one-particle contraction of the $n$-particle 
correlation function:
\begin{equation}
c (\mathbf{x}) = \sum_{j=1}^{N} \int_{\Omega^{N-1}} \rho_N (\Lambda_j) \, \prod_{k=1,k\neq j}^N d \mathbf{x}_k, \label{conc_def}
\end{equation}
which, in thermal equilibrium, can be expressed in terms of the Boltzmann distribution:
\begin{eqnarray}
\nonumber
\rho_N (\mathbf{x}_1,\mathbf{x}_2,\cdots,\mathbf{x}_N) \hspace{4cm}  \\
\nonumber
= Z (s)^{-1} \exp\left(-\frac{\psi(\mathbf{x}_1,\mathbf{x}_2,\cdots,\mathbf{x}_N)}{k_B T}\right),\\
\label{equil_rho_n}
\end{eqnarray}
where $ T $ is the temperature, $ k_B $ is Boltzmann's constant, and the partition function
for the solute particles confined by a surface $ \partial \Omega (s) $ is given by:
\begin{eqnarray}
\nonumber
Z(s) &=& \int_{\Omega^N} \exp\left(-\frac{\psi(\mathbf{x}_1,\mathbf{x}_2,\cdots,\mathbf{x}_N)}{k_B T}\right) \, \prod_{k=1}^N d \mathbf{x}_k \\
\nonumber
& = & \int_{\Omega^N} \exp\left(-
\frac{V(\mathbf{x}_1,\mathbf{x}_2,\cdots,\mathbf{x}_N)}{k_B{T}}\right)\cdot \\
&&\exp\left(\frac{-\sum_{j=1}^N \Phi (\mathbf{x}_j,s)}{k_B T}\right)  
\prod_{k=1}^N d \mathbf{x}_k.
\label{Z_spherical_R}
\end{eqnarray}
Substituting~Eq.~(\ref{conc_def})~and~Eq.~(\ref{wall_force}) into~Eq.~(\ref{P_wl_conc}), 
we obtain for the osmotic wall pressure:
\begin{eqnarray}
\nonumber
P_{\mathrm{wl}} & = & \frac{1}{|\partial \Omega|}\int_{\Omega} h(\mathbf{x}) c(\mathbf{x}) \, d\mathbf{x} \\
\nonumber
&=&  \frac{1}{|\partial \Omega|}\int_{\Omega} - \frac{\partial \Phi (\mathbf{x},s)}{\partial s}\cdot \\
\nonumber
&&\cdot
\sum_{j=1}^{N} \int_{\Omega^{N-1}} \rho_N (\Lambda_j) \, \left(\prod_{k=1,k\neq j}^N d \mathbf{x}_k\right) d \mathbf{x} \\
\nonumber
& = & \sum_{j=1}^N \frac{1}{|\partial \Omega|}\int_{\Omega^N} - \frac{\partial \Phi (\mathbf{x_j},s)}{\partial s}  \cdot \\
\nonumber
&&\cdot
\rho_N (\mathbf{x}_1,\mathbf{x}_2,\cdots,\mathbf{x}_N) \, \prod_{k=1}^N d \mathbf{x}_k \\
  & = &  
  \frac{k_B{T}}{|\partial \Omega|} \frac{\partial }{\partial s} \left. \ln \left(Z(s)\right) \right|_{s=s_0}
 \label{P_wl_general}.   
\end{eqnarray}
and in the last expression, we have indicated that the 
pressure is to be calculated for the initially 
specified chamber wall $ s=s_0 $.
Our result~Eq.~(\ref{P_wl_general}) agrees with the thermodynamic 
definition of pressure as the (functional) derivative of 
the Helmholtz free energy $ - k_B T \ln Z $ with respect 
to volume $ V $ along our one-parameter family of
deformations, for which $ d V = |\partial \Omega| \, ds $.
We remark that since the bounding walls are assumed to 
be impermeable to the solute, the confinement energy 
must diverge at the boundary: 
$\Phi(\mathbf{x},s) \rightarrow \infty$ as $\mathbf{x} \rightarrow \partial \Omega (s)$,
which has the effect of restricting integration to only the interior of 
the chamber.  The reason that the osmotic wall pressure is only equal to the volume derivative of the free energy along a certain one-parameter family of shape 
deformations appears to be related to the fact that nonlocal interactions (such as by a soft wall potential) make the \emph{local} pressure no longer constant within a container, and therefore along the surface, so that different kinds of shape deformations will have different free energy costs per volume of deformation.  

Next we develop a statistical mechanical representation for the
osmotic fluid pressure~Eq.~(\ref{P_fl_definition}), 
our second notion of osmotic 
pressure.  We shall again assume that
the system at statistical equilibrium has Boltzmann statistics~Eq.~(\ref{equil_rho_n}), 
from which it follows that the expression for $ \nabla \bar{p} (\mathbf{x}) $ 
in~Eq.~(\ref{def_avg_pressure_grad})
can be simplified as follows:
\begin{eqnarray}
\nonumber
\nabla \bar{p} (\mathbf{x}) & = & 
\frac{k_B T}{Z} \nabla_{\mathbf{x}} \sum_{j=1}^{N}
\int_{\Omega^{N-1}} \exp \left(- \frac{\Psi (\Lambda_j)}{k_B T}\right) \prod_{k=1,k\neq j}^N d \mathbf{x}_k \\
& = &  k_B T \nabla c (\mathbf{x}).
\label{P_fl_general}
\end{eqnarray}
From~Eq.~(\ref{P_fl_definition}) and~Eq.~(\ref{P_fl_general}), we obtain
\begin{equation}
P_{\text{fl}} (\mathbf{x}) 
 = k_B T c (\mathbf{x}). \label{P_fl_integrate}
\end{equation}
This shows that our form of the osmotic fluid pressure connects
with macroscopic notions in nonequilibrium thermodynamics 
\citep{katchalsky1965}.

\subsection{Hard-Wall Limit}
\label{sec:hardwall}
We now show how these formulas relate to the van't Hoff law for 
osmotic pressure both on the confining wall and in the fluid
in the case that the particles are noninteracting and the 
confining potential has a classical 
hard-wall potential of the form:
\begin{eqnarray}
\Phi(\mathbf{x},s) = \left\{  
\begin{array}{ll}
0,      & \mathbf{x} \in \Omega (s) \\
\infty, & \mathbf{x} \not\in \Omega (s)
\end{array}
\right.
.
\label{Phi_hard_wall}
\end{eqnarray}
Substitution of this ``hard-walled" 
confining potential into~Eq.~(\ref{Z_spherical_R}), 
we have that $ Z(s) $ is just the $N$th power of volume $ V(s) $  of the chamber $ \Omega (s) $.
By~Eq.~(\ref{P_wl_general}), the osmotic wall pressure is then:
\begin{eqnarray}
P_{\mathrm{wl}} = k_B {T} c_0
\end{eqnarray}
where $c_0 = {N}/{V(s_0)}$
is the concentration, recovering the well-known 
van't Hoff's Law 
equation \citep{vanthoff1887}. Similarly, from~Eq.~(\ref{P_fl_integrate}),
we have:
\begin{eqnarray}
P_{\mathrm{fl}}(\mathbf{x}) = \frac{N k_B T (V(s_0))^{N-1}}{Z (s_0)} = 
\frac{N k_B T}{V (s_0)} =  k_B T c_0,
\label{P_fl_hard_wall_2}
\end{eqnarray}
and again find that
the classical van't Hoff's Law is 
recovered in the fluid pressure.

We remark that the key to 
obtaining the classical 
van't Hoff's Law in both cases 
was to consider the hard-walled
limit of the confining potentials
which arises naturally when 
the length scale of 
particle-wall interactions are very
small relative to the diameter of the 
confining chamber.  In this regime 
a theory similar to our approach 
was developed in~\citep{brenner1996},
in which a careful analysis is made of
the balance of mechanical forces arising
from the solute interactions with the 
walls, again under the assumption 
that the chamber diameter is much 
larger than the length-scale of the 
particle interaction forces.
While this assumption typically holds for 
macroscopic systems, when the chamber 
size becomes sufficiently small, 
the results of this limit are no 
longer strictly valid and corrections are required to 
the classical theory reflecting some of the
microscopic features of the system.  

\subsection{\label{sec:steric_interaction}
Steric Interactions with the Chamber Walls of 
Non-negligible Length-Scale} 
For microscopic systems, the length-scale 
on which the solute particles interact with 
the wall may be non-negligible relative to
to the diameter of the chamber.  For example, 
if a rigid spherical particle of radius $\ell$ 
is confined in a spherical chamber having radius $R$,
the steric interactions would confine the particle 
center to have only configurations with 
$|\mathbf{x}| < R - \ell$.  We shall now illustrate
how this effects the osmotic pressure when 
$\ell$ is comparable 
in size to $R$.

More precisely, in the case of a 
spherical chamber of radius $R$ 
the confinement potential is of the general form:
\begin{eqnarray}
\Phi(\mathbf{x},R) 
   & = & \int_{\partial \Omega(R)} \phi(|\mathbf{y} - \mathbf{x}|)\frac{R_0^2}{R^2} d\mathbf{y},
\end{eqnarray}
where $\phi(\rho)$ denotes the solute-wall potential for a given reference
sphere of radius $R_0$ and 
$\mbox{det} \left( \nabla \Psi^{-1}_s(\mathbf{y})\right) = {R_0^2}/{R^2}$.
Now for $R_{\star} = R - \ell < R$ we have:
\begin{eqnarray}
\phi(\rho) = \left\{  
\begin{array}{ll}
0,      & \rho \geq \ell \\
\infty, & \rho < \ell 
\end{array}
\right. 
\end{eqnarray}
where $\rho = |\mathbf{y} - \mathbf{x}|$.
By radial symmetry this gives 
for $r = |\mathbf{x}|$ the solute 
confinement potential:
\begin{eqnarray}
\Phi(r,R) = \left\{  
\begin{array}{ll}
0,      & r \leq R - \ell \\
\infty, & r > R - \ell
\end{array}
\right.
.
\end{eqnarray}

The partition function and its derivative are
obtained explicitly as:
\begin{eqnarray}
Z(R) & = & \frac{4\pi}{3}(R - \ell)^3 \\
\frac{\partial Z(R)}{\partial R} 
& = & 4\pi (R - \ell)^2 .
\end{eqnarray}
From~Eq.~(\ref{P_fl_definition}), 
the osmotic fluid pressure is given by:
\begin{eqnarray}
P_{\mathrm{fl}}(\mathbf{0}) & = & \frac{k_B{T}}{\frac{4\pi}{3}(R - \ell)^3}.
\end{eqnarray}
From~Eq.~(\ref{P_wl_general}), the osmotic-wall pressure is given by:
\begin{eqnarray}
P_{\mathrm{wl}} & = & \frac{k_B T}{Z} = 
\frac{k_B{T}}{\frac{4\pi}{3}(R - \ell)^3}
\left(1 - \frac{\ell}{R}\right)^2 
\end{eqnarray}
where $\ell < R$.

The difference in these two notions of pressure in this case
can be interpreted geometrically.  In particular, the 
confinement forces restrict the solute particles within a 
spherical region of radius $R_{\star} = R - \ell$.  This has the effect 
of generating within the 
confinement region the same local confinement forces as a 
hard-walled potential with a wall occupying the spherical 
shell of radius $R_{\star} = R - \ell$ and yields the same 
osmotic-fluid pressure as in the hard-walled case.  
However, the particle-wall interactions, which generate 
the confinement forces, occur from a wall occupying the 
spherical shell of radius $R$, as opposed to a spherical
shell of radius $R_{\star}$ as would occur in the strictly 
"hard-wall" case.  From
the principle of equal and opposite forces, the particle-wall
interactions now exert forces in the normal direction 
over a wall with greater surface area than in the short-range
hard-walled case, 
 thus reducing the osmotic wall pressure by 
the geometric factor $\left(1 - {\ell}/{R}\right)^2 
= \left({R_{\star}}/{R}\right)^2$, which is
the ratio between the surface areas.  

While this example is rather special, it illustrates 
clearly one mechanism by which differences 
with the classical theory of osmotic pressure can arise. 
This also  illustrates how notions of osmotic pressure 
in terms of the wall pressure and fluid pressure can 
differ markedly when 
the length scales of the particle-wall 
interactions become comparable to the diameter of the 
chamber.  We next 
 demonstrate how osmotic pressures 
behave for smooth potentials having long-range 
particle-wall interactions.

\section{\label{sec:softwall_interaction}
Osmotic Pressure for Non-interacting Solute Particles 
Confined by a Soft-Walled Potential}

We now examine the behavior of the osmotic pressure for a system
of non-interacting solute particles which 
interact with the chamber wall through a smooth long-range
potential. For simplicity, we will consider the osmotic pressure exerted by a single
solute particle; the case of $N $ non-interacting solute particles of course simply multiplies the pressures by a factor of $ N$.  We shall consider here the class of repulsive potentials 
for a spherical chamber of radius $R_0$ 
of the form:
\begin{eqnarray}
\phi(\rho) & = & C\rho^{-\alpha},
\end{eqnarray}
which, under volume dilations to new radii $ R $, 
 induce  effective wall potentials 
\begin{eqnarray}
\Phi (r,R) &= \frac{2 \pi C R_0^2}{R^{\alpha}}
\frac{
(1 - r)^{2 - \alpha} - (1+r)^{2-\alpha}
}{(\alpha - 2)r}
\end{eqnarray}
We consider only $\alpha > 2$ to ensure that 
$\lim_{r\rightarrow R} \Phi(r,R) = \infty$.

The partition 
function and its derivative can be expressed as:
\begin{eqnarray}
\nonumber
Z(R) & = & 4\pi R^3 \int_0^1 
\exp\left(
-\frac{
(1 - r)^{2 - \alpha} - (1+r)^{2-\alpha}
}{(\alpha - 2){\lambda}r}
\right)
r^{2} dr \\
\end{eqnarray}
and 
\begin{eqnarray}
\nonumber
\frac{\partial Z}{\partial R}
& = & 
\frac{3Z(R)}{R} \\
\nonumber
& + & 
\frac{4\pi R^2}{{\lambda}}
\left(
\frac{\alpha}{\alpha - 2}
\right)
\int_0^1
\left(
\frac{
(1 - r)^{2 - \alpha} - (1+r)^{2-\alpha}
}{r}
\right) \\
\nonumber
&&
\hspace{1.5cm}
\exp\left(
-\frac{
(1 - r)^{2 - \alpha} - (1+r)^{2-\alpha}
}{(\alpha - 2){\lambda}r}
\right)
r^{2} dr
\\
\end{eqnarray}
where $\lambda = k_BTR^{\alpha}/2 \pi C R_0^2$
can be regarded as a reduced temperature. 
These expressions were obtained by 
using~Eq.~(\ref{def_Phi_int}) and~Eq.~(\ref{Z_spherical_R}) with $ R=s$ 
and 
making the change of variable $r = Rr'$.
This gives the osmotic fluid pressure:
\begin{eqnarray}
P_{\mathrm{fl}}(\mathbf{0})
& = & 
\frac{k_B{T}}{\frac{4\pi}{3}R^3}
Q_1(\lambda,\alpha)
\end{eqnarray}
and the osmotic wall pressure:
\begin{eqnarray}
P_{\mathrm{wl}}
& = & 
\frac{k_B{T}}{\frac{4\pi}{3}R^3}
Q_2(\lambda,\alpha)
\end{eqnarray}
where
\begin{eqnarray}
\nonumber
Q_1(\lambda,\alpha) = 
\exp\left({-\frac{2}{\lambda}}\right) \cdot
\hspace{4cm}
 \\
 \cdot
\left[
3
\int_0^1
\exp\left(
-\frac{
(1 - r)^{2 - \alpha} - (1+r)^{2-\alpha}
}{(\alpha - 2){\lambda}r}
\right)
r^{2} dr
\right]^{-1}
&&
\end{eqnarray}
and
\begin{eqnarray}
&&Q_2(\lambda,\alpha) 
=  
1 +  
\left(
\frac{1}{3\lambda}
\right)
\left(
\frac{\alpha}{\alpha - 2}
\right)
\cdot\\
\nonumber
&&
\hspace{0.5cm}
\cdot
\frac{
\int_0^1
\left(
\frac{
(1 - r)^{2 - \alpha} - (1+r)^{2-\alpha}
}{r}
\right)
\exp\left(
-\frac{
(1 - r)^{2 - \alpha} - (1+r)^{2-\alpha}
}{(\alpha - 2){\lambda}r}
\right)
r^{2} dr
}
{
\int_0^1
\exp\left(
-\frac{
(1 - r)^{2 - \alpha} - (1+r)^{2-\alpha}
}{(\alpha - 2){\lambda}r}
\right)
r^{2} dr  
}.
\end{eqnarray}
For this system with the power-law potential, we can express $Q_1$ 
compactly as
\begin{eqnarray}
Q_1 = \frac{\frac{4}{3}\pi R^3}{Z(R)}
\exp\left(-\frac{2}{\lambda}\right)
\end{eqnarray}
and $Q_2$ as
\begin{eqnarray}
\nonumber
Q_2(\lambda,\alpha) 
& = & 
1 + 
\frac{\alpha}{3}
\left\langle
\frac{
(1 - r)^{2 - \alpha} - (1+r)^{2-\alpha}
}{(\alpha - 2)\lambda r}
\right\rangle \\
&=& 1 + \frac{\alpha \langle \Phi \rangle}{3 k_B T}.  
\end{eqnarray}
The last expression shows 
 that $Q_2$ is one plus a
term proportional to the
ensemble average of the energy of the system
over all configurations of the confined particle with respect to the Boltzmann distribution. 

\begin{figure}
\centering
\epsfxsize = 3.5in
\epsffile[37 192 545 585]{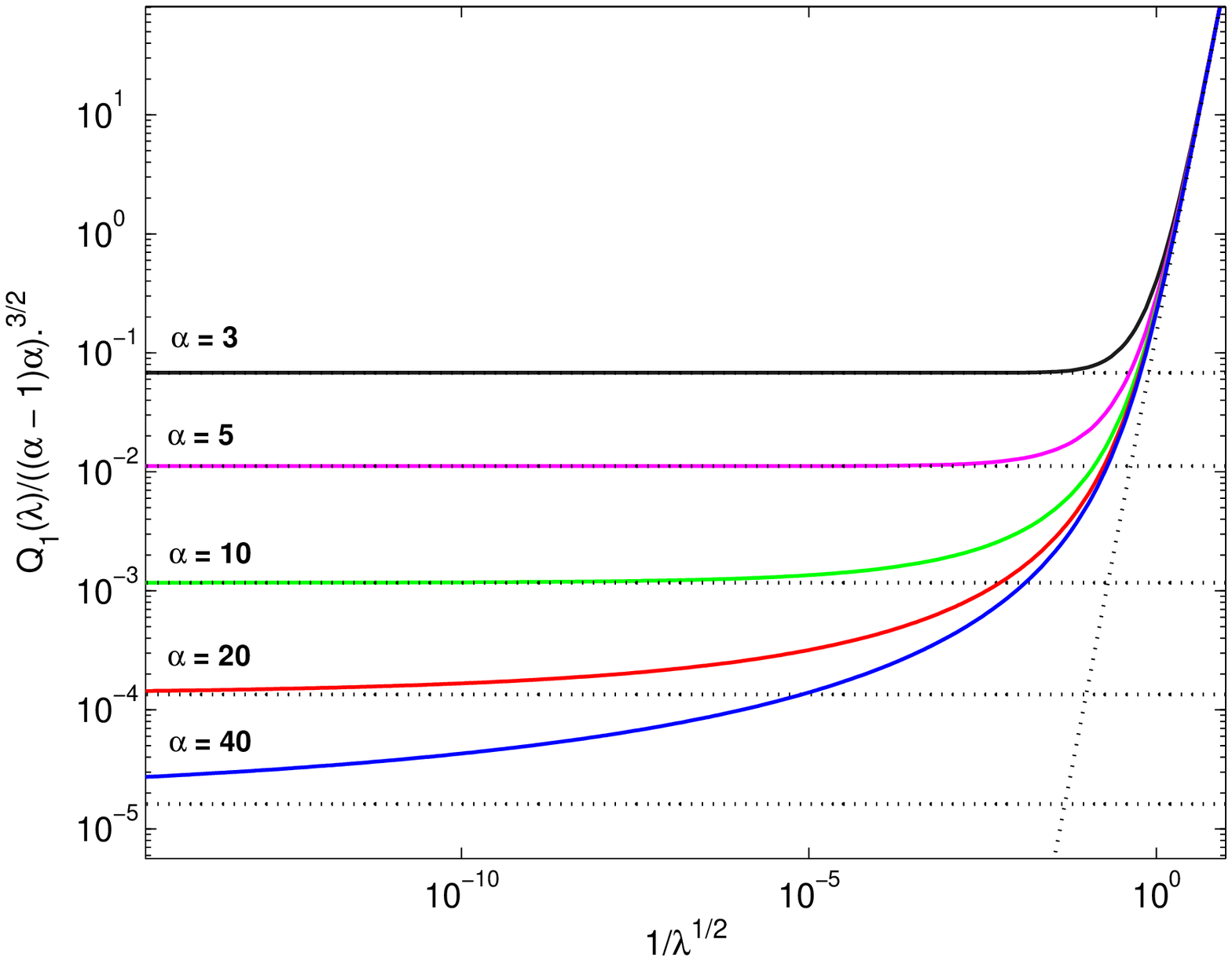}
\caption[]{$Q_1$ Correction Factor for Soft-Walled Potential.
For $\lambda \rightarrow \infty$ we have $1/\lambda^{1/2} \rightarrow 0$
and $Q_1 \rightarrow 1$.  This corresponds to the plotted horizontal lines.
For $\lambda \rightarrow 0$
we have
$1/\lambda^{1/2} \rightarrow \infty$ and
${Q_1}/{((\alpha - 1)\alpha)^{3/2}} \sim 
\left(
\frac{9\sqrt{3\pi}}{4}
\lambda^{3/2} 
\right)^{-1}$.
This corresponds to the plotted diagonal line on the right.
\label{figure_Q_1_corrections}
}
\end{figure}

\begin{figure}
\centering
\epsfxsize = 3.5in
\epsffile[37 192 545 585]{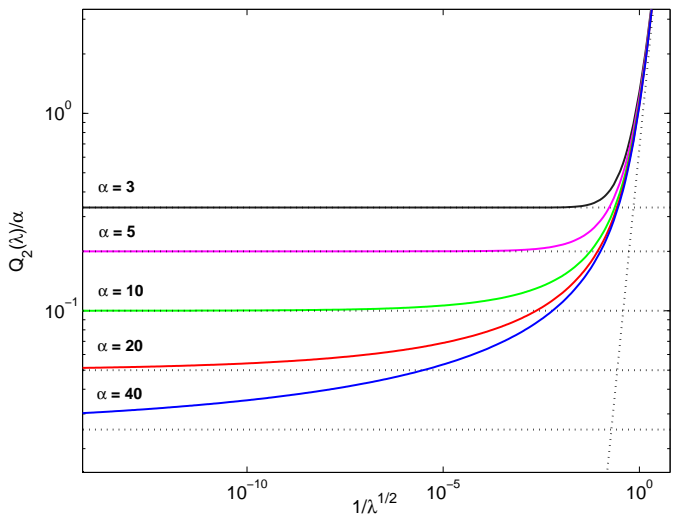}
\caption[]{$Q_2$ Correction Factor for Soft-Walled Potential.
For $\lambda \rightarrow \infty$ 
we have $1/\lambda^{1/2} \rightarrow 0$ and
$Q_2 \rightarrow 1$.
This corresponds to the plotted horizontal lines.
For $\lambda \rightarrow 0$ 
we have 
$1/\lambda^{1/2} \rightarrow \infty$
and
$Q_2/\alpha \sim {2}/{3\lambda}$.
This corresponds to the plotted diagonal line on the right.
\label{figure_Q_2_corrections}
}
\end{figure}

We observe first of all that the soft-walled nature of the potential produces corrections to the van't Hoff law for the osmotic pressure of the chamber in both the osmotic wall pressure and fluid pressure.  To compare these correction factors with each other, we first examine their asymptotic behaviors in the limits $ \lambda \rightarrow 0 $ and $ \lambda \rightarrow \infty $.  
Note that the term $\lambda$ represents the 
ratio of the energy scale of the 
thermal fluctuations relative
to that of the confining 
potential, and therefore these limits correspond to low temperature and high temperature limits, respectively.   As  $\lambda \rightarrow \infty$ 
we have $Q_1, Q_2 \rightarrow 1$,
which recovers the hard-walled limit.  
On the other hand, 
for small $\lambda$, 
$Q_1 \sim 
\left(
\frac{9\sqrt{3\pi}}{4}
\left(
\frac{\lambda}{(\alpha - 1)\alpha}
\right)^{3/2}
\right)^{-1}$
and 
$Q_2 \sim
{2\alpha}/{3\lambda}$, showing 
a behavior similar to a system at low temperature 
in which the structure of the long-range particle-wall 
interactions  plays 
 a significant role.  We note that the correction factors for the osmotic fluid pressure and osmotic wall pressure diverge at different rates with respect to $ \lambda $ as $ \lambda \rightarrow 0 $, indicating that the osmotic fluid and wall pressures will deviate significantly both from each other and from the van't Hoff law for soft-walled confining potentials with energy scales large compared to the temperature.  On the other hand, plots of the behavior of the correction factors as a function of $ 1/\lambda^{1/2} $ (see Figures~\ref{figure_Q_1_corrections} 
and~\ref{figure_Q_2_corrections}) show qualitatively similar behavior, in that the ratio of the osmotic pressures to the values given by the van't Hoff law increases monotonically from $1$ as $ 1/\lambda^{1/2} $ increases and diverges as $ \lambda \rightarrow 0 $.

\section{\label{sec:interact_string}
Osmotic Pressure of Solute Dimers 
Connected by ``Strings" 
and Confined 
by a Hard-Walled Potential}

We shall now discuss how the length scale of interaction between the
solute particles can affect the osmotic pressures when this length scale is comparable to the chamber size.  
To illustrate through analytical formulas the basic
mechanism by which this occurs, we 
shall consider the rather special 
case of two solute particles which 
are connected by a ``string" 
and confined by a ``hard-walled" spherical potential.
More precisely, the potential energy of the system is given by
\begin{eqnarray}
\Psi(\mathbf{x}_1,\mathbf{x}_2,R) 
= \Phi(\mathbf{x}_1,R) 
+ 
\Phi(\mathbf{x}_2,R)
+ 
V(\mathbf{x}_1,\mathbf{x}_2) 
\end{eqnarray}
where $\Phi$ is the hard-wall potential given in~Eq.~(\ref{Phi_hard_wall}).
The potential $V$ models the 
interaction between the two solute 
particles, located at 
$\mathbf{x}_1$ and $\mathbf{x}_2$, 
and is given by:
\begin{eqnarray}
V(\mathbf{x}_1,\mathbf{x}_2)
= \left\{
\begin{array}{ll}
0, & \mbox{ if $|\mathbf{x}_2 - \mathbf{x}_1| < \ell$} \\
\infty, & \mbox{otherwise} \\
\end{array}
\right. .
\end{eqnarray}
This potential can be given the physical
interpretation of 
no coupling 
between the two particles until they attempt to 
separate past a distance $\ell$, at which point
an infinitely strong restoring force constrains 
the particles to be a distance less than or equal 
$\ell$.  This form of coupling is analogous to 
connecting the two particles by an inelastic 
``piece of string" of length $\ell$. 
The osmotic pressure for $M$
noninteracting strings would of course just multiply the
1-string pressures we calculate by $ M$.  

\begin{figure}
\centering
\epsfxsize = 3.0in
\epsffile[1 1 356 267]{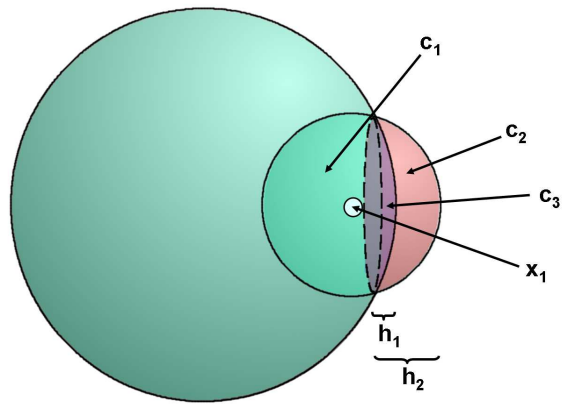}
\caption[]{Geometry of the Intersection of Two Spheres.
The surface of any two spheres intersect in a 
circle carried in a common plane (dashed circle).  
We shall refer to this as the ``plane of intersection" 
and further assume that the center of the smaller
sphere lies within the volume of the larger sphere.
For any such plane we may define
three volumes by ``slicing" the small and large 
sphere along this plane.  We let $c_1$ 
denote the volume the sliced part of the larger 
sphere lying to the left of the ``plane 
 of 
intersection".  We let $c_2$ denote the volume 
associated with the sliced part of the small
sphere lying to the right of the ``plane 
 of 
intersection".  We let $c_3$ denote 
the volume associated with the sliced part 
of the larger sphere
lying to the right of the ``plane 
of 
intersection.'' 
  To characterize the 
numerical values of the volume of 
these regions we define $h_1$ to be 
height of the spherical cap $c_3$
and $h_2$ the height of the spherical 
cap $c_2$.
\label{figure_sphere_intersect1}
}
\end{figure}
Using this form of the potentials we have:
\begin{eqnarray}
\Psi(\mathbf{x}_1,\mathbf{x}_2,R) = \left\{  
\begin{array}{ll}
0,      & (\mathbf{x}_1,\mathbf{x}_2) \in \Gamma(R,\ell) \\
\infty, & (\mathbf{x}_1,\mathbf{x}_2) \not \in \Gamma(R,\ell) 
\end{array}
\right. 
\end{eqnarray}
where 
\begin{eqnarray}
\nonumber
\Gamma(R,\ell) = \left\{(\mathbf{x}_1,\mathbf{x}_2) | 
                        |\mathbf{x}_1| < R,
                        |\mathbf{x}_2| < R,
                        |\mathbf{x}_2 - \mathbf{x}_1| < \ell
                 \right\} \\
\end{eqnarray}
with $\Gamma$ the admissible region in $\mathbb{R}^3 \times \mathbb{R}^3$
in which both of the solute particles are a distance
no more than $\ell$ apart and contained within
the spherical chamber of radius $R$.
In this case, the partition function
can be expressed in terms of a few quantities having a 
straightforward geometric interpretation and
computed exactly.  In particular,
the partition function
\begin{eqnarray}
Z(R) & = & \int_{\Omega^2} 
\exp\left(
-\frac{\Psi(\mathbf{x}_1,\mathbf{x}_2,R)}{k_B{T}}
\right)
d\mathbf{x}_1\, 
d\mathbf{x}_2
\end{eqnarray}
 is the volume of the region
$\Gamma(R,\ell) \subset \mathbb{R}^6$.  

To compute this volume, 
we shall find it convenient to split the configuration space
into two parts: 
$\Gamma^{(1)}(R,\ell) = \{|\mathbf{x}_1| < R - \ell \}$
and $\Gamma^{(2)}(R,\ell) = \{R - \ell < |\mathbf{x}_1| < R \}$,
where it is to be understood that $\Gamma^{(1)},\Gamma^{(2)} \subset \Gamma$. 
In the first region, for each $\mathbf{x}_1$ the second particle $\mathbf{x}_2$
is free to assume values within 
 the entire sphere of radius $\ell$ about $\mathbf{x}_1$
having volume $\frac{4\pi}{3}\ell^3$.  In the second region, for each $\mathbf{x}_1$
the second particle $\mathbf{x}_2$ must lie within the intersection of
the sphere of radius $\ell$ centered at $\mathbf{x}_1$
and the sphere of radius $R$ centered at $\mathbf{0}$.   The volume of the region so defined will be denoted by $W_{\star}(|\mathbf{x}_1|)$. 
Using this decomposition, the partition function can be expressed as:
\begin{eqnarray}
\nonumber
Z(R) & = & \int_{\Gamma^{(1)}} \frac{4\pi}{3}\ell^3 d\mathbf{x}_1
        +  \int_{\Gamma^{(2)}} W_{\star}(|\mathbf{x}_1|) \, 
         d\mathbf{x}_1 \\
         \nonumber
     & = & \left(\frac{4\pi}{3}\ell^3\right)\left(\frac{4\pi}{3} (R - \ell)^3\right) \\    
     & + & 4\pi\int_{R - \ell}^{R} W_{\star}(r) 
     r^2 dr 
\end{eqnarray}
where the volume $W_{\star}$ can be expressed as:
\begin{eqnarray}
W_{\star}(r) 
& = & C(R,h_1(r)) + \frac{4\pi}{3}\ell^3 - C(\ell,h_2(r))
\end{eqnarray}
with $C$ denoting the volume of a spherical cap of height $h$ for a
sphere of radius $r$ (see figure \ref{figure_sphere_intersect1}):
\begin{eqnarray}
C(r,h) & = & \frac{1}{3} \pi h^2 (3r - h).
\end{eqnarray}
The $h_1(r) = R - u_0(r)$ is the height of the spherical 
cap $c_3$ 
and $h_2(r) = r + \ell - u_0(r)$ is the height
of the spherical cap $c_2$, 
 where 
$u_0(r) = (r^2 - \ell^2 + R^2)/{2r}$ is the radius
corresponding to the plane which contains  the 
intersection of the two spheres (Figure~\ref{figure_sphere_intersect1}).

From this the exact solution for the partition function and 
its derivative can be computed to obtain:
\begin{eqnarray}
Z(R,\ell) & = & 
\frac {16}{9}{\pi}^{2}{R}^{3}{\ell}^{3} 
- {\pi}^{2}{R}^{2}{\ell}^{4} 
+ \frac{1}{18}{\pi}^{2}{\ell}^{6}
\end{eqnarray}
and
\begin{eqnarray}
\frac{\partial Z}{\partial R}(R,\ell) & = & 
\frac{16}{3}{\pi}^{2}{\ell}^{3}{R}^{2}
- 2{\pi}^{2}R{\ell}^{4}.
\end{eqnarray}
The osmotic pressures are then given
for $\ell < R$ by:
\begin{eqnarray}
P_{\mathrm{fl}}(\mathbf{0}) & = & 
\frac{2k_B{T}}
{
\frac{4}{3}\pi {R}^{3}
}
\left(
\frac{1}{
1 - \frac{9}{16}\left(\frac{\ell}{R}\right) + \frac{1}{32}\left(\frac{\ell}{R}\right)^3 
}
\right)\\
P_{\mathrm{wl}} & = & 
\frac{k_B{T}}{\frac{4}{3}\pi R^{3}}
\left(
\frac{
1 - \frac{3}{8}\left(\frac{\ell}{R}\right)
}
{
1 - \frac{9}{16}\left(\frac{\ell}{R}\right) + \frac{1}{32}\left(\frac{\ell}{R}\right)^3 
}
\right).
\end{eqnarray}
This gives the correction to the classical theory 
when the interactions of the solute particles 
becomes non-negligible relative to the diameter
of the chamber.  For $ \ell > 2 R $, the particle interaction becomes trivial and both osmotic wall and fluid  pressure assume the van't Hoff~\citep{vanthoff1887} law value of $ 2 k_B T/\frac{4}{3} \pi R^3 $.  

The above expressions 
verify that in the limit 
$\ell \rightarrow 0$, the osmotic wall pressure 
converges to the classical van't Hoff 
pressure $ k_B T/\frac{4}{3} \pi R^3 $ of a single confined particle.  This is to be expected as the two-particle string should coalesce into a single entity in the tight-binding limit.  A similar transition between van't Hoff law behavior for the osmotic wall pressure exerted by dimers in the weak-binding and tight-binding limits was shown for spring-like dimers in  numerical simulations with the stochastic immersed boundary method in~\citep{atzberger2005ib}.

The osmotic fluid pressure, by contrast, approaches the value $ 2 k_B T/\frac{4}{3} \pi R^3 $ as $ \ell \rightarrow 0 $, which corresponds to the van't Hoff law for two particles in the sphere, even though the two particles are tightly bound.  This provides another example for how the pressure built up in the fluid near the center need not be directly linked in magnitude to the pressure exerted by the solute particles on the chamber wall.

\section{\label{sec:interact_n_gen}
Corrections to van't Hoff Law for Interacting Particles in
 Spherical Chambers with Soft-Walled Potentials}

We shall now consider the more general case 
of $N$ interacting solute particles confined 
in a spherical chamber of radius $R$ with a 
soft-walled potential of the form:
\begin{eqnarray}
\nonumber
\Psi(\mathbf{x}_1,\cdots,\mathbf{x}_N,R)
& = & 
V(\mathbf{x}_1,\cdots,\mathbf{x}_N)
+
\sum_{k = 1}^N 
\Phi(\mathbf{x}_k,R) \\
\end{eqnarray}
where $V(\mathbf{x}_1,\cdots,\mathbf{x}_N)$
models the interactions between the 
particles and $\Phi(\mathbf{x}_k,R)$ 
confines the $k^{th}$ solute particle to 
the interior of the chamber, with
$\Phi \rightarrow \infty$ as 
$|\mathbf{x}_k| \rightarrow R$.

From~Eq.~(\ref{Z_spherical_R}) and~Eq.~(\ref{P_wl_general}), the osmotic pressure 
can be expressed as:
\begin{eqnarray}
\label{P_wl_n_interacting}
\nonumber
P_{\mathrm{wl}} 
& = & 
\frac{N k_B{T}}{\frac{4}{3}\pi R^3} \\
\nonumber
& + & 
\frac{1}{4\pi R^2}
\left(
\sum_{k = 1}^N
\left\langle
-\nabla_{\mathbf{x}_k}
V(\mathbf{x}_1,\mathbf{x}_2,\ldots,\mathbf{x}_N) 
\cdot
\mathbf{x}_k
\right\rangle
\right) \\
\nonumber
& + & 
\frac{1}{4\pi R^2}
\left(
\sum_{k = 1}^N
\left\langle
-\nabla_{\mathbf{x}_k}
\Phi(\mathbf{x}_k,R) 
\cdot
\mathbf{x}_k
-
\frac{\partial \Phi(\mathbf{x}_k,R)}{\partial R}
\right\rangle
\right)  \\
\end{eqnarray}
where $\mathbf{x} = (\mathbf{x}_1,\cdots,\mathbf{x}_N)$
and $\langle\rangle$ denotes the ensemble average over
the particle configurations weighted by the Boltzmann factor 
$\exp\left({-{\Psi}/{k_B{T}}}\right)$.
This was derived by making the change of variable
$\mathbf{x} = R\mathbf{x}'$ in $Z$ and differentiating
in $R=s$. 

This expression can be used to characterize 
the corrections to the classical theory 
and the relative contributions of the 
microscopic effects of the system.
The first term is the classical pressure
one would expect from the van't Hoff's law.
The second term arises from the 
particle-particle interactions, and
the third term is from the particle-wall interactions. 
It is important to note that while we have 
described the source of each distinct term, 
the terms in fact are coupled by the ensemble 
average which depends on the combination of 
these effects.

The second correction term has an intuitive 
interpretation as follows.  Since the contribution
to the pressure is of the form
$-\nabla V \cdot \mathbf{x}$ we have that when 
the particle interaction force acting on any 
particle is toward the chamber center, 
the
force acting on the bounding wall will be 
relieved and a negative contribution will be 
made to the pressure.
This suggests one mechanism by which polymerization 
reduces osmotic pressure relative to 
 free 
monomers.  For a polymer in a typical configuration, 
most of the monomers will, for entropic reasons,
be outside a boundary layer of the wall.  As
a consequence, any individual monomer that
makes an excursion toward the wall, would
on average experience a pulling force 
toward the chamber center. From the
second correction term this will 
reduce the osmotic pressure relative to 
free monomers.  This perspective also provides another way of interpreting the wall pressure results for the string model from Section~\ref{sec:interact_string}.  

The third term gives the corrections
that arise 
 from long-range interactions
of the solute particles with the boundary
wall.  In the hard-walled limit 
this term approaches zero.   

We observe that the correction terms to the van't Hoff law in~Eq.~(\ref{P_wl_n_interacting}),
other than the partial derivative with respect to chamber radius $ R$, 
take the form of the virial from classical mechanics.
Virial expansions of the pressure can be found in several
statistical mechanical textbooks~\citep{IBler:mcsp} to describe departures of a dilute
gas or solution from an ideal
noninteracting particle limit.    These results are however generally developed within
the context of a large macroscopic chamber, whereas our focus is on systems for which
interaction length scales become comparable to those of the chamber.  

\section{\label{sec:confined_polymers}
Osmotic Pressure of Confined Polymers}
\begin{figure}
\centering
\epsfxsize = 3.5in
\epsffile[0 0 359 296]{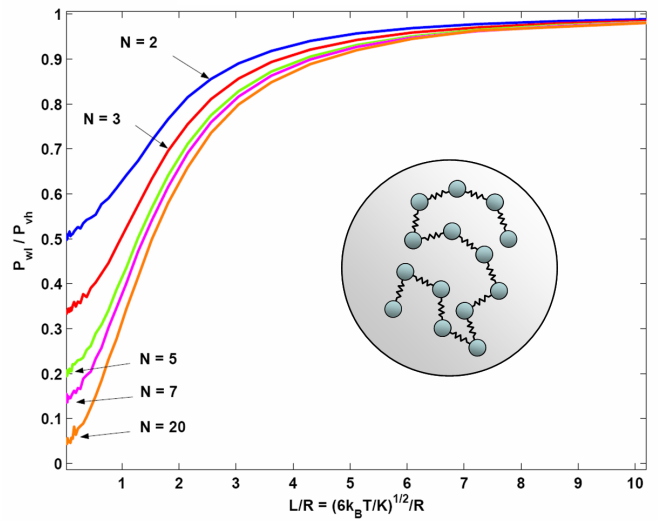}
\caption[]{
Osmotic Pressure of Confined Polymer.
Correction Factor for a Linear Polymer Confined in a
Spherical Chamber.  The classical van't Hoff's Law for $N$
free monomers is $P_{vh} = {Nk_B{T}}/{\frac{4}{3}\pi R^3}$.
For stretching an individual bond between monomers, we let
$L = ({6k_B{T}}/{K})^{1/2}$ denote the length at which the 
bond energy becomes $3 k_B{T}$.
\label{figure_linear_polymer_corrections}
}
\end{figure}

To illustrate how this theory can be applied in 
practice, we now present some numerical results 
for a model of a polymer chain of $N$ monomers
confined in a spherical chamber with a hard-wall
confining potential.  The monomers of the polymer 
will have coupling given by the harmonic 
bonding energy:
\begin{eqnarray}
V(\mathbf{x}_1,\cdots,\mathbf{x}_N)
& = & \sum_{j = 2}^{N} 
\frac{K}{2}
\left|\mathbf{x}_{j} -
\mathbf{x}_{j - 1}\right|^2.
\end{eqnarray}

The osmotic pressure of the system can be computed
from~Eq.~(\ref{P_wl_n_interacting}), which in this case
reduces to:
\begin{eqnarray}
P_{\mathrm{wl}}
& = & 
\frac{Nk_B{T}}{\frac{4}{3}\pi R^3}
+
\frac{1}{4\pi R^2}
\sum_{k = 1}^n
\left\langle
-\nabla_{\mathbf{x}_k}
V\cdot \mathbf{x}_k
\right\rangle . 
\end{eqnarray}

We will focus on how the osmotic pressure behaves as the bonding strength $ K$ is varied so that
 the length-scale 
$L = \sqrt{6k_B{T}/K}$ of bond fluctuations varies between lengths small and large relative to the chamber size. 
 For very small $K \ll 6k_B{T}/R^2$ the fluctuations in
the bond length are expected to be very large 
and the $N$ monomers to behave independently.
For very large $K \gg 6k_B{T}/R^2$ the fluctuations in the bond 
length are expected to be very small and the 
$N$ monomers to behave similarly to a single 
particle.  Thus in the extreme cases the classical
osmotic pressures are expected, corresponding to an $N $ particle
or single particle system. 

To obtain the osmotic wall pressure for intermediate values of $ K$, 
the correction factors were estimated numerically using the
Monte-Carlo method with Metropolis Sampling~\citep{landau2000}; see Figure~\ref{figure_linear_polymer_corrections}.  
We 
find that as the bond stiffness $K \rightarrow 0$, 
 the osmotic 
wall pressure approaches the classical van't Hoff osmotic pressure for 
$N$ free monomers.  As $K \rightarrow \infty$ we find that
the osmotic wall pressure approaches the 
classical van't Hoff osmotic pressure for a single particle.
The numerical results show how the theoretical framework 
 can be used to 
capture the regime relevant to microscopic chambers, such 
as a polymer confined in a vesicle, in which the bond length is
comparable to the chamber size.  
As can be seen from  Figure~\ref{figure_linear_polymer_corrections}, 
the transition between 
the extremes in the bonding strength is gradual and occurs 
smoothly in $K$. 
A similar numerical study for the osmotic pressure of polymers was conducted within the stochastic immersed boundary method in~\citep{atzberger2005ib}.

\section{\label{sec_conclusion}Conclusion}
We have shown that the osmotic pressure 
deviates markedly from classical macroscopic theories 
when the length-scale of the system becomes
sufficiently small so that the chamber diameter is
comparable in size to the length-scale of the
interactions of the solute particles.  In particular,
we have explored two ways in which this can occur, 
either through interactions among the solute molecules
or interactions of the solute molecules with the wall.  
While the classical formulas are not directly applicable in this 
regime, we have shown that a theoretical framework 
for the equilibrium osmotic pressures can still be 
developed provided additional microscopic 
features of the system are taken into 
account.  We considered how the osmotic pressure is
manifested in terms of forces exerted on the confining wall 
and in terms of the hydrostatic pressure of the fluid in the 
interior of the chamber.  We showed through several examples 
how these notions of osmotic pressure can differ for microscopic 
chambers.  In particular, we showed how the osmotic wall 
pressure of a two particle string and an N-particle polymer
interpolate between the van't Hoff laws associated with 
the extremes of strong and weak bonding, which give effectively
a single particulate entity or many individual particulate entities.  

While the osmotic fluid pressure is related in a simple local manner to the osmotic pressure 
defined in nonequilibrium thermodynamics~\citep{katchalsky1965,IBler:mcsp}, it behaves a bit less intuitively.
This was demonstrated for the two particle string, where the tight binding limit produced an osmotic fluid 
pressure corresponding to two particulate entities.  This is in contrast to the osmotic wall pressure 
which reflected a single confined entity.  The distinction appears to be related to the local nature of 
the osmotic fluid pressure, 
which can vary throughout the chamber and in particular differ between the wall region and center of the 
chamber.  Fluid pressure is only expected to be constant in mechanical equilibrium when the solute-wall forces 
in the system are local, but here we have been considering interactions between particles and walls on length scales 
comparable to the chamber size.  An important consequence is that the fluid pressure need not be constant 
throughout the chamber interior and the pressure felt by the wall need not be equal to the 
build-up in fluid pressure in the chamber center. 

The theoretical framework presented here 
should be readily applicable to the study of 
equilibrium osmotic phenomena in many microscopic 
physical systems arising in 
biology and technological
applications.  While only 
equilibrium systems where considered here,
it may be possible to use many of the central 
ideas of this theory to ultimately formulate
a non-equilibrium (or near-equilibrium) theory 
for microscopic osmotic phenomena. 

\ifx\undefined\allcaps\def\allcaps#1{#1}\fi\newcommand{\noopsort}[1]{}
  \newcommand{\printfirst}[2]{#1} \newcommand{\singleletter}[1]{#1}
  \newcommand{\switchargs}[2]{#2#1}
  \ifx\undefined\allcaps\def\allcaps#1{#1}\fi\def\cprime{$'$}

\begin{acknowledgments}
The author P.J.A. was supported by NSF VIGRE Postdoctoral Research
Fellowship DMS - 9983646 and NSF Mathematical Biology Grant DMS - 0635535.
The author P.R.K. was partially supported by NSF CAREER Grant
DMS-0449717. 
We would like to thank Charles Peskin, George Oster, 
Philip Pincus, and Frank Brown for helpful suggestions.
We also thank the reviewers for helpful comments.
\end{acknowledgments}

\appendix

\section{\label{sec:inversion} 
Obtaining the Particle-Wall Interaction
Force and Potential from the Confinement 
Force and Potential (Inversion Formulas)}

From~Eq.~(\ref{P_wl_definition}) we see 
that in order to obtain explicit expressions 
for the osmotic pressure it is useful to 
have an expression for
$\mathbf{G}(|\mathbf{x} - \mathbf{y}|)$.
In modeling systems, we may readily have only 
the
confining force $\mathbf{F}_{\text{wl}} (\mathbf{x})$ 
while the
detailed particle-wall interaction
forces must somehow be inferred.  
For radial symmetric potentials of the form
$\mathbf{F}_{\text{wl}} (\mathbf{x})
 = F(|\mathbf{x}|)\frac{\mathbf{x}}{|\mathbf{x}|}$
and
$\mathbf{G}(\mathbf{r}) = 
G(|\mathbf{r}|) \frac{\mathbf{r}}{|\mathbf{r}|} $, 
this requires solving the following inverse problem for $ G $ given $ F$:
\begin{eqnarray}
-F(|\mathbf{x}|)\frac{\mathbf{x}}{|\mathbf{x}|}
& = & 
\int_{\partial \Omega} G(|\mathbf{y} - \mathbf{x}|) 
\frac{\mathbf{x} - \mathbf{y}}{|\mathbf{x} - \mathbf{y}|}
\, d \mathbf{y}. 
\end{eqnarray}
By dotting both sides with $\frac{\mathbf{x}}{|\mathbf{x}|}$, this becomes
the scalar problem:
\begin{eqnarray}
-F(|\mathbf{x}|)
& = & 
\int_{\partial \Omega} G(|\mathbf{y} - \mathbf{x}|) 
\frac{\mathbf{x} - \mathbf{y}}{|\mathbf{x} - \mathbf{y}|}\cdot
\frac{\mathbf{x}}{|\mathbf{x}|}\, d \mathbf{y}.
\end{eqnarray}
In spherical coordinates, the integral transform can 
be expressed as:
\begin{eqnarray}
-F(r) & = & \frac{-\pi R}{r^2} \int_{R - r}^{R + r} 
G(\rho) (r^2 - R^2 + \rho^2) d\rho 
\end{eqnarray}
where $\rho = |\mathbf{x} - \mathbf{y}|$ and 
$r = |\mathbf{x}|$.

To ensure a unique solution to the inverse problem
we shall make the assumption that the particle-wall
interactions occur only over a distance less than 
the radius $R$ from the chamber wall, that is 
$g(\rho) = 0$ for $\rho \geq R$.  Under this 
assumption the transform can be inverted exactly with 
the solution formula:
\begin{eqnarray}
\nonumber
g(\rho) &=& \left(
\frac{1}{2\pi R \rho^2}
\right)
\left(
(R+\rho) F (R - \rho) \right. \\
\nonumber
& & \qquad \left. + \rho (R-\rho) F^{\prime} (R-\rho)
+
\int_0^{R - \rho}
F(s) \,
ds
\right) \\
\end{eqnarray}
from which $G(\rho)$  
is readily obtained.

Alternatively, for 
 a given wall 
 potential $\Phi(\mathbf{x},R)$ 
the inverse problem is to determine a
$\phi(\rho)$ so that for all $\mathbf{x}$ 
\begin{eqnarray}
\Phi(\mathbf{x},R) & = & 
\int_{|\mathbf{y}| = R}  
\phi(|\mathbf{x} - \mathbf{y}|) d\mathbf{y}.
\end{eqnarray}
In the case that the potential $\Phi$ is radially 
 symmetric 
in $r = |\mathbf{x}|$, a change of variable allows
for the integral to be expressed as
\begin{eqnarray}
\Phi(r,R) & = & \frac{2\pi R}{r}\int_{R - r}^{R + r}
\phi(\rho)\rho d\rho.
\end{eqnarray}
A unique solution can be found for this problem
under the assumption 
 that $\phi(\rho) = 0$ for $\rho \geq R$.
By differentiating both sides in $r$ 
and substituting 
 $\rho = R - r$, 
the following inversion formula is obtained:
\begin{eqnarray}
\nonumber
\phi(\rho) & = & 
\frac{1}{2\pi R \rho}  
\left(
\Phi(R - \rho,R)
+
(R - \rho) \frac{\partial \Phi}{\partial r} 
(R - \rho,R)
\right). \\
\end{eqnarray}


\begin{thebibliography}{0}
\expandafter\ifx\csname natexlab\endcsname\relax\def\natexlab#1{#1}\fi
\expandafter\ifx\csname bibnamefont\endcsname\relax
  \def\bibnamefont#1{#1}\fi
\expandafter\ifx\csname bibfnamefont\endcsname\relax
  \def\bibfnamefont#1{#1}\fi
\expandafter\ifx\csname citenamefont\endcsname\relax
  \def\citenamefont#1{#1}\fi
\expandafter\ifx\csname url\endcsname\relax
  \def\url#1{\texttt{#1}}\fi
\expandafter\ifx\csname urlprefix\endcsname\relax\def\urlprefix{URL }\fi
\providecommand{\bibinfo}[2]{#2}
\providecommand{\eprint}[2][]{\url{#2}}

\end{thebibliography}


\begin{thebibliography}{10}


\bibitem{atzbergerPump2006}
{\sc P.~A.~Atzberger and C.~S.~Peskin},
{\em A Microfluidic Pump Exploiting Solute 
Diffusion and Osmotic Effects},
(in preparation).

\bibitem{bazant2004}
{\sc M. Z. Bazant and T. M. Squires},
{\em Induced-Charge Electrokinetic Phenomena: Theory and Microfluidic Applications},
Phys. Rev. Let., vol. 92, no. 6, February (2004).

\bibitem{nardi1999}
{\sc J.~Nardi, R.~Bruinsma, and E.~Sackmann},
{\em Vesicles as Osmotic Motors},
Phys. Rev. Let., Vol. 82. Num. 25, (1999).

\bibitem{brenner1996}
{\sc D.~C.~Guell and H.~Brenner},
{\em Physical Mechanism of Membrane Osmotic Phenomena},
Ind. Eng. Chem. Res.,35, 3004-3014, (1996).

\bibitem{su2002}
{\sc Y. C. Su, L. Lin, and A. P. Pisano},
{\em A Water-Powered Osmotic Microactuator},
Journal of Microelectromechanical Systems,
vol. 11, no. 6, December (2002).

\bibitem{theeuwes1976}
{\sc F. Theeuwes and S. I. Yum},
{\em Principles of the Design and Operation of Generic Osmotic
Pumps for the Delivery of Semisolid or Liquid Drug Formulations},
Annals of Biomedical Engineering,
vol. 4, no. 4, December (1976).

\bibitem{verma2002}
{\sc R. K. Verma, D. M. Krishna, and D. Garg},
{\em Formulation Aspects in the Development of Osmotically Controlled 
Oral Drug Deliver Systems},
Journal of Controlled Release, 79, pp.~7--27, (2002).

\bibitem{wolgemutha_hydra_2004}
{\sc C.~W.~Wolgemutha, A.~Mogilner, and G.~Oster},
{\em The hydration dynamics of polyelectrolyte gels 
with applications to cell motility and drug delivery},
Eur Biophys J, 33: 146–158, (2004).

\bibitem{hammel1999}
{\sc H.~T.~Hammel},
{\em Evolving ideas about osmosis and capillary fluid exchange}
The FASEB Journal, vol. 13, February (1999).

\bibitem{duquette2001}
{\sc P. P. Duquette, P. Bissonnette, and J. Y. Lapointe},
{\em Local Osmotic Gradients Drive the Water Flux Associated 
with Na+/glucode Cotransport},
PNAS, vol. 98, no. 7, pp.~3796--3801, March (2001).

\bibitem{spring1999}
{\sc K. R. Spring},{\em Epithelial Fluid Transport -- A Century of Investigation},
News Physiol. Sci.
vol. 4, June (1999).

\bibitem{hoppensteadt2001}
{\sc F. C. Hoppensteadt and C. S. Peskin},
{\em Modeling and Simulation in Medicine and the Life Sciences},
Springer-Verlag, New York (2002).

\bibitem{go:doff}
{\sc G.~Oster and C.~S. Peskin}, {\em Dynamics of osmotic fluid flow}, in
  Swelling Mechanics: {F}rom Clays to Living Cells and Tissues, T.~Karalis,
  ed., Springer-Verlag, Berlin, 1992, pp.~731--742.

\bibitem{douzou1994}
{\sc P. Douzou},
{\em Osmotic Regulation of Gene Action},
Proc. Natl. Acad. Sci, USA
vol. 91, pp.~1657--1661, March (1994).

\bibitem{evilevitch2003}
{\sc A.~Evilevitch, L.~Lavelle, C.~M.~Knobler, E.~Raspaud and W.~M.~Gelbart},
{\em Osmotic pressure inhibition of DNA ejection from
phage.}
Proc. Natl. Acad. Sci. USA 100, 9292–9295 (2003). 

\bibitem{thecell}
{\sc B.~Alberts, A.~Johnson, J.~Lewis, M.~Raff, K.~Roberts, and P.~Walker},
{\em Molecular Biology of the Cell}, Garland Publishing, (2002).

\bibitem{wolgemutha_myx_2004}
{\sc C.~W.~Wolgemutha and G.~Oster},
{\em The Junctional Pore Complex and the
Propulsion of Bacterial Cells},
J Mol Microbiol Biotechnol, 7:72–77 (2004).

\bibitem{block2003}
{\sc K.~C.~Neuman and S.~M.~Block},
{\em Optical trapping}, Review of Scientific Instruments,
    volume 75, number 9, (2004).

\bibitem{danuser2003}
{\em Computational analysis of f-actin turnover in cortical actin meshworks using 
fluorescent speckle microscopy},
{\sc A.~Ponti,  P.~Vallotton,  W.~Salmon, C.~Waterman-Storer, and G.~Danuser},
Biophys. J., 84 :3336-3352, (2003).

\bibitem{einstein1926}
{\sc A.~Einstein}, {\em Investigations on the Theory of the Brownian Movement},
  Dover Publications, New-York (1926).

\bibitem{lippincott2003}
{\em Development and Use of Fluorescent Protein
Markers in Living Cells},
{\sc J.~Lippincott-Schwartz and G.~H.~Patterson},
Science vol. 300, 4, April (2003).

\bibitem{levich1962}
{\sc V.~G.~Levich},
{\em Physico-Chemical Hydrodynamics},
Englewood Cliffs NJ: Prentice Hall, (1962).

\bibitem{lipowsky1999}
{\sc R. Lipowsky},
{\em From Membranes to Membrane Machines 
:In Statistical Mechanics of Biocomplexity},
Springer-Verlag, Berlin, pp. 1--32, (1999).

\bibitem{vanthoff1887}
{\sc J.H. van `t Hoff},
{\em The Role of Osmotic Pressure in the Analogy Between Solutions and Gases},
Zeitschrift fur physikalische Chemie,
vol. 1, pp. 481-508, (1887)

\bibitem{atzberger2005ib}
{\sc P.~J. Atzberger, P,~R. Kramer, and C.~S. Peskin}, 
{\em A Stochastic Immersed Boundary Method for 
Fluid-Structure Interactions at Microscopic Length Scales},
(to appear in J. of Comp. Phys. doi 10.1016/j.jcp.2006.11.015), 
(2006).

\bibitem{jfb:sd}
{\sc J.~F. Brady and G.~Bossis}, {\em Stokesian dynamics}, in Annual review of
  fluid mechanics, vol.~20 of Annu. Rev. Fluid Mech., Annual Reviews, Palo
  Alto, CA, pp.~111--157, (1988).

\bibitem{katchalsky1965}
{\sc A.~Katchalsky and P.~F.~Curran},
{\em Nonequilibrium Thermodynamics in Biophysics},
Ch 10, Harvard University Press, Cambridge, Massachusetts, (1965).

\bibitem{IBler:mcsp}
{\sc L.~E. Reichl}, {\em A modern course in statistical physics}, John Wiley \&
  Sons Inc., New York, second~ed., (1998).

\bibitem{grosberg2002}
{\em The physics of charge inversion in chemical and biological
systems},
{\sc A.~Y.~Grosberg, T.~T.~Nguyen, and B.~I.~Shklovskii},
Reviews of Modern Physics, Vol. 74, April, (2002).

\bibitem{prabhu2005}
{\sc V.~M.~Prabhu},
{\em Counterion structure and dynamics in polyelectrolyte solutions},
Current Opinion in Colloid \& Interface Science,10, (2005).

\bibitem{prabhu2005}
{\sc V.~M.~Prabhu},
{\em Counterion structure and dynamics in polyelectrolyte solutions},
Current Opinion in Colloid \& Interface Science,10, (2005).

\bibitem{IBrk:sp2}
{\sc R.~Kubo, M.~Toda, and N.~Hashitsume}, {\em Statistical physics. {I}{I}},
  Springer-Verlag, Berlin, second~ed., section~4.
\newblock Nonequilibrium statistical mechanics, (1991).

\bibitem{IBldl:ctp9}
{\sc L.~D. Landau and E.~M. Lifshitz}, {\em Course of theoretical physics.
  {V}ol. 9: {S}tatistical physics}, Pergamon Press, Oxford, 1980, ch.~IX.

\bibitem{auroy1992}
{\sc P.~Auroy, Y.~Mir, and L.~Auvray},
{\em Local Structure and Density Profile of Polymer Brushes},
Phys. Rev. Lett. ,Vol. 69, Num. 1 (1992).

\bibitem{IBldl:ctp6}
{\sc L.~D. Landau and E.~M. Lifshitz}, {\em Course of theoretical physics.
  {V}ol. 6: {F}luid Mechanics}, Butterworth-Heinemann, Oxford, 1987, ch.~II.

\bibitem{peskin2002}
{\sc C.~S.~Peskin}, {\em The immersed boundary method}, 
Acta Numerica, 11, pp.~1--39, (2002).

\bibitem{groot1951}
{\sc S.~R.~de~Groot},
{\em Thermodynamics of Irreversible Processes},
(North-Holland, Amsterdam, 1951), Chap. 45.

\bibitem{landau2000}
{\sc D.~P.~Landau and K.~Binder},
{\em A Guide to Monte Carlo Simulations in Statistical Physics},
Cambridge University Press, New York, NY (2000).

\end{thebibliography}
\end{document}